\documentclass[]{article}  
\usepackage{graphicx}
\usepackage{amsfonts}
\usepackage{amssymb}
\usepackage{latexsym}

\newenvironment{pf}{\unskip{\bf Proof:}}{\unskip{\hfill $\Box$}}

\newcommand{\lemlab}[1]{\label{lemma:#1}}
\newcommand{\theolab}[1]{\label{theo:#1}}

\newcommand{\figlab}[1]{\label{fig:#1}}
\newcommand{\seclab}[1]{\label{section:#1}}

\newcommand{\theoref}[1]{\ref{theo:#1}}

\newcommand{\figref}[1]{\ref{fig:#1}}
\newcommand{\eqref}[1]{(\ref{eq:#1})}
\newcommand{\secref}[1]{\ref{section:#1}}

\newtheorem{theorem}{Theorem}
\newtheorem{lemma}[theorem]{Lemma}
\newtheorem{cor}[theorem]{Corollary}
\newtheorem{conj}{Conjecture}[section]

{\catcode`\@=11
\gdef\setft#1#2#3{%
\def\@oddfoot{
{\setbox0=\hbox{#1}
\setbox1=\hbox{#3}
\ifdim\wd0>\wd1
\dimen0=\wd0
\box0\hfil#2\hfil\hbox to\dimen0{\hfil\hfil\box1}
\else \dimen0=\wd1
\hbox to\dimen0{\box0\hfil }\hfil#2\hfil\box1 \fi
}}} }

\def\complaint#1{}
\def\withcomplaints{
\newcounter{mycomplaints}
\def\complaint##1{\refstepcounter{mycomplaints}%
\ifhmode%
\unskip%
{\dimen1=\baselineskip \divide\dimen1 by 2 %
\raise\dimen1\llap{\tiny -\themycomplaints-}}\fi%
\marginpar{\tiny [\themycomplaints]: ##1}}%
}

\def\P{{\cal P}}
\def\R{\mathbb{R}}
\def\bD{{\partial D}}

\newlength\abovesectionskip
\newlength\belowsectionskip
\makeatletter
\def\section{\@startsection{section}{1}{\z@}{-\abovesectionskip}%
               {\belowsectionskip}{\normalfont\Large\bfseries}}
\makeatother

\begin{document}

\title{{\bf Vertex-Unfoldings\\of Simplicial Polyhedra}}

\author{
Erik~D.~Demaine%
\thanks{Dept. of Computer Science, Univ. of Waterloo,
Waterloo, Ontario N2L 3G1, Canada.
eddemaine@\allowbreak uwaterloo.ca.}
\and
David~Eppstein%
\thanks{Dept. of Information and Computer Science, 
Univ. of California, Irvine CA
92697-3425, USA.  
eppstein@\allowbreak ics.uci.edu.
Supported by NSF grant CCR-9912338.}
\and
Jeff~Erickson%
\thanks{Dept. of Computer Science, Univ. of Il\-linois at
Ur\-bana-Cham\-paign; 
http://\allowbreak www.\allowbreak cs.\allowbreak uiuc.\allowbreak
edu/\~{}jeffe; jeffe@\allowbreak cs.\allowbreak uiuc.\allowbreak
edu.
Partially supported by a Sloan Fellowship and NSF CAREER award
CCR-0093348.}
\and
George~W.~Hart%
\thanks{http://\allowbreak www.georgehart.com/;
george@\allowbreak georgehart.com.}
\and
Joseph~O'Rourke%
\thanks{
Dept. of Computer Science, Smith Col\-lege, North\-ampton,
MA 01063, USA.
orourke@\allowbreak cs.smith.edu.
Supported by NSF grant CCR-9731804.
}
}


\maketitle

\begin{abstract} 
We present two algorithms for unfolding the surface of any
polyhedron, all of whose faces are triangles, 
to a nonoverlapping, connected planar layout.
The surface is cut only along polyhedron edges.
The layout is connected, but it may have a
disconnected interior:
the triangles are connected at vertices, but not necessarily
joined along edges.  
\end{abstract}
 
\section{Introduction}
\seclab{Introduction} 

It is a long-standing open problem to decide whether 
every convex polyhedron may be cut along edges and unfolded
flat in one piece without overlap, i.e., unfolded to
a simple polygon.
This type of unfolding has been termed \emph{edge-unfolding};
the unfolding consists of the faces of the polyhedron joined
along edges.
In contrast,
unfolding via arbitrary cuts easily
leads to nonoverlap.
See~\cite{o-fucg-00} for history and
applications to manufacturing.
Recently it was established that not every nonconvex polyhedron
may be edge-unfolded, even if the polyhedron is \emph{simplicial},
that is, all of its faces are triangles~\cite{bdek-up-99,bdems-uptf-01}.
In this paper we loosen the meaning of ``in one piece'' to
permit a nonoverlapping connected region
that (in general) does not form a simple
polygon, because its interior is disconnected.
We call such an unfolding a \emph{vertex-unfolding}:
the faces of the polyhedron are joined at vertices (and sometimes edges).
With this easier goal we obtain a positive result:
the surface of
every simplicial polyhedron, convex or nonconvex, of any genus,
may be cut along edges and unfolded to a planar,
nonoverlapping, connected layout.
Our proof relies on the restriction that every face is a triangle.
The problem remains open for nonsimplicial polyhedra with
simply connected faces (see Section~\secref{Discussion}).

\section{Overview of Algorithm}
\seclab{Overview}
Let $\P$ be a simplicial polyhedron,
and let $G$ be the \emph{lattice graph} of the face lattice of the
polyhedron:
the nodes of $G$ are the facets (triangles),
edges, and vertices of $\P$, with an arc for
each incidence.

Define a \emph{facet path} in $G$ to be a path that 
alternates between vertices and facets, includes
each facet exactly once, 
and never repeats the same vertex twice in a row.
In such a path $p = ( \ldots, v_1, f, v_2, \ldots )$,
$v_1$ and $v_2$ are distinct vertices of $f$.

Our first observation is that if $G$ contains a facet path $p$,
then a vertex-unfolding exists.
For the triangle $f$ can be placed
inside a vertical strip with $v_1$ and $v_2$ on the left and right
strip boundaries.  Doing this for each triangle in $p$
yields a nonoverlapping unfolding, connected at the strip
boundaries.

However, we do not know whether every lattice graph
has a facet path.  We can prove this only for polyhedra of genus
zero (Theorem~\theoref{facet.cycle} in Section~\secref{facet.cycle}).
\begin{figure}[htbp]
\centering
\includegraphics[width=0.5\linewidth]{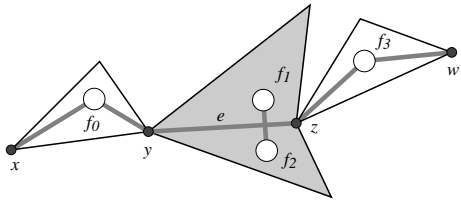}
\caption{An unfolding path.}
\figlab{TQ-tree}
\end{figure}
Instead we establish that that every $G$
(for any simplicial polyhedron, of any genus)
has an ``unfolding path,'' which is roughly a facet path 
that may also include quadrilaterals.
More precisely, an \emph{unfolding path} is a path in $G$
that alternates between vertices and nonvertices,
covers each facet exactly once,
and never repeats the same vertex twice in a row.
An edge-node of a path covers the two adjacent facets;
otherwise a facet is covered if it is a node of the path.
An example is shown in Fig.~\figref{TQ-tree}.
Here the path is $(x, f_0, y, e, z, f_3, w)$,
with $e$ covering $f_1$ and $f_2$.
Of course every facet path is an unfolding path.

As is evident in the figure, if the quadrilateral $Q=f_1 \cup f_2$
is nonconvex, it is no longer as straightforward to place $Q$
inside a vertical strip.  However, we can always choose
to open up the end of $e=yz$ at which
$Q$ has a convex angle ($y$ in the figure), which then allows
$Q$ to be placed in a strip.
See Fig.~\figref{equilateral}.
\begin{figure}[htbp]
\centering
\includegraphics[width=0.75\linewidth]{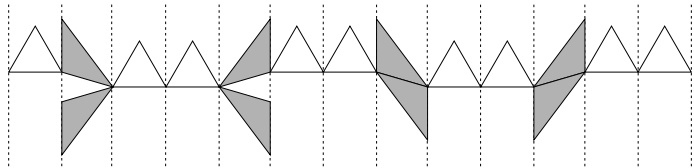}
\caption{Laying out an unfolding path in vertical strips.
The quadrilaterals are shaded.
(The example is contrived.)}
\figlab{equilateral}
\end{figure}

Therefore, once an unfolding path is found for $G$,
a vertex-unfolding can be achieved.

\section{Algorithm for Arbitrary Genus}
\seclab{Algorithm}
Our proof that every lattice graph $G$ for a simplicial polyhedron $\P$
has an unfolding path is via an algorithm
that finds such a path.
The first step is to convert the surface into a 
simplicial complex $D$ forming a topological
disk (henceforth, a \emph{disk})
by cutting a sufficient number of polyhedron edges.
For a polyhedron of genus $0$, just one edge needs be cut.
The algorithm operates on $D$, finding a path through its
triangles, and occasionally employing a quadrilateral
when it can no longer extend with a triangle.
In particular, the unfolding path starts at a vertex $s$
on the boundary of $D$, and ends at a boundary vertex $t \neq s$.

Let $D$ be a disk containing
at least one triangle.
We'll let $\bD$ represent its boundary.
Let $s$ and $t$ be distinct boundary vertices.
Vertex $s$ has two distinct neighbors $s_1$ and $s_2$ on $\bD$,
and $t$ has two distinct neighbors $t_1$ and $t_2$ on $\bD$.
Call the triangles incident to $s s_i$ and $t t_i$, $i=1,2$,
\emph{$s$-wings} and \emph{$t$-wings} respectively. 
Although there may be as many as four distinct wings,
there could be as few as one, because several of the wings might coincide.
Say that a triangle $T$ \emph{breaks the disk $D$} if $D \setminus T$ is
not a topological disk.
Define an $s$-wing to be a \emph{good wing} if
it is not incident to $t$ and does not break the disk;
similarly a $t$-wing is good if it is not incident
to $s$ and does not break the disk.
Good wings permit easy advancement of the facet path.
Throughout we let $\pi(s,D,t)$ represent an unfolding path from
$s$ to $t$ through $D$, and use $\oplus$ to represent path
concatenation: 
$\pi(a,D_1,b) \oplus \pi(b,D_2,c)$
is the path from $a$ to $b$ in $D_1$ joined at $b$ to the
path from $b$ to $c$ in $D_2$:
$( a , \ldots, b, \ldots, c )$.

\begin{enumerate}

\item There is a good wing $T$.
Let $T=\triangle s s_1 u$; all other cases are symmetric.
We join $(s, T, u)$ to the recursively constructed
path in the remainder (Fig.~\figref{Cases}(1)):
$$\pi(s,D,t) = (s, T, u) \oplus \pi( u, D \setminus T, t )\;.$$
Note that $u \in  D \setminus T$ and $u \neq t$, which justifies
the recursion.
Whether $T$ is an $s$- or a $t$-wing, the structurally similar
path construction suffices to reduce to a smaller disk,
either from the $s$- or from the $t$-end.

\begin{figure}[htbp]
\centering
\includegraphics[width=0.75\linewidth]{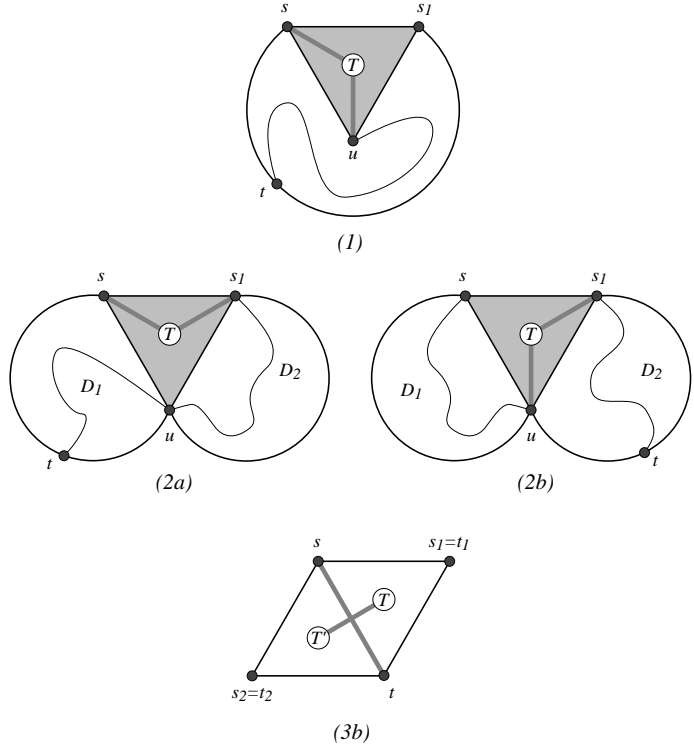}
\caption{Cases of the algorithm.}
\figlab{Cases}
\end{figure}

\item There is an $s$-wing $T$ that is not incident to $t$,
or a $t$-wing that is not incident to $s$.
Again let $T=\triangle s s_1 u$; all other cases are symmetric.
Because $T$ is not good, it must break the disk, which implies
that $u$ is on $\bD$.
Let $D_1$ and $D_2$ be the disks separated by $T$,
with $s \in D_1$. Note that neither $s_1$ nor $u$ can be $t$.
\begin{enumerate}
\item $t \in D_1$ (Fig.~\figref{Cases}(2a)):
$$\pi(s,D,t) = 
(s, T, s_1) \oplus \pi( s_1, D_2, u ) \oplus \pi( u, D_1, t)\;.$$
\item $t \in D_2$ (Fig.~\figref{Cases}(2b)):
$$\pi(s,D,t) = 
\pi( s, D_1, u ) \oplus (u, T, s_1) \oplus \pi( s_1, D_2, t)\;.$$
\end{enumerate}

\item Every $s$-wing is incident to $t$, and every $t$-wing is incident
to $s$.
Then it must be that $D$ is either a single triangle,
or a quadrilateral.
\begin{enumerate}
\item $D$ is a triangle $T$.  Then $(s, T, t)$ is an unfolding path
for $D$.
\item $D$ is a quadrilateral $T \cup T'$; Fig.~\figref{Cases}(3b).
Then $(s, st, t)$  is an unfolding path
for $D$.

\end{enumerate}
\end{enumerate}

It is not difficult to
see that the algorithm can be
implemented to require time only
linear in the number of triangles in the
disk $D$.

\section{Algorithm Proof}
\seclab{Proof}

\begin{theorem}
Any triangulated topological disk $D$
has an unfolding path connecting any two
distinct boundary vertices
$s$ and $t$.
\theolab{disk}
\end{theorem}
\begin{pf}
The proof is by induction on the number of triangles $n$ in $D$.
It obviously holds for $n=1$.
The algorithm just described clearly results in an unfolding path
for $D$, by construction.
The only issues that remains are verifying that the cases indeed
exhaustively cover the possibilities,
and that in each case, the conditions of the induction hypothesis 
hold.

A good wing by definition neither breaks the disk, nor is incident
to $t$ (resp. $s$) for an $s$- (resp. $t$-) wing.
Case~1 covers good wings, and Cases~2 and~3 cover the two ways
to fail being a good wing:  Case~2 for wings that do not violate
the incidence condition (in which case they must violate the
breaking condition), and Case~3 the wings that do violate the
incidence condition.
Thus the cases are
mutually exclusive
and comprehensive.  

That the induction hypothesis holds in each case is
easily seen.  We are careful to ensure that the start
and end vertices in each recursive application are distinct
boundary vertices, and that the subcomplex being traversed
is a disk.

The only issue that remains is why Case~3 requires $D$ to be
a triangle or a quadrilateral.
Suppose there are two $s$-wings, so that
$s_1$ and $s_2$ are distinct.  Each must be incident to $t$,
so $t$ is neither $s_1$ nor $s_2$.
Thus the two wings must be
$T=\triangle s s_1 t$ and $T'= \triangle s s_2 t$.
If $s_1 t$ or $s_2 t$ is not a boundary edge,
then some wing of $t$ is not incident to $s$, which would place
us in Case~2.  So both are boundary edges, and $D$ is the quadrilateral
$T \cup T'$.

Finally, suppose that there is just one $s$-wing $T=\triangle s s_1 u$.
Then both $s s_1$ and $s u$ must be boundary edges,
with either $s_1 = t$ or $u = t$.
In either case, if $s_1 u$ is not a boundary edge, there would be
a $t$-wing not incident to $s$, again leading to Case~2.
Because we know Case~2 does not hold,  $s_1 u$ must be a boundary
edge, and $D = T$.
\end{pf}

By our remarks in Section~\secref{Overview},
Theorem~\theoref{disk} suffices to establish our main result:

\begin{theorem}
The surface of any simplicial polyhedron $P$ (of any genus)
may be 
vertex-unfolded (in linear time): cut along
edges and unfolded to a nonoverlapping, connected
planar layout.
In the layout,
adjacent vertical strips each contain
one or two triangles of $P$.
\theolab{any.genus}
\end{theorem}
\noindent

Note that the resulting unfolding could be viewed as
a \emph{hinged dissection}~\cite{f-dpf-97} of the surface;
see, for example, Fig.~\figref{cube}.
\begin{figure}[htbp]
\centering
\includegraphics[width=0.7\linewidth]{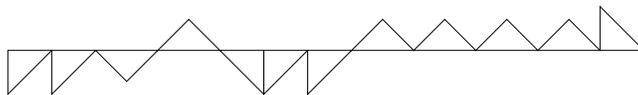}
\caption{Unfolding of the surface of a triangulated cube.}
\figlab{cube}
\end{figure}

\section{Examples}
\seclab{Examples}
We have implemented 
the algorithm of Section~\secref{Algorithm} and applied it to
a number of convex polyhedra.
Fig.~\figref{vunf.all} shows several examples.
The polyhedra were generated as convex hulls of randomly generated
points.
Most unfoldings were in fact facet paths:
encountering a quadrilateral (in Case~3b of the algorithm)
was somewhat rare.
However, the figure illustrates only cases in which one or
more quadrilaterals occur.
\begin{figure}[htbp]
\centering
\includegraphics[width=\linewidth]{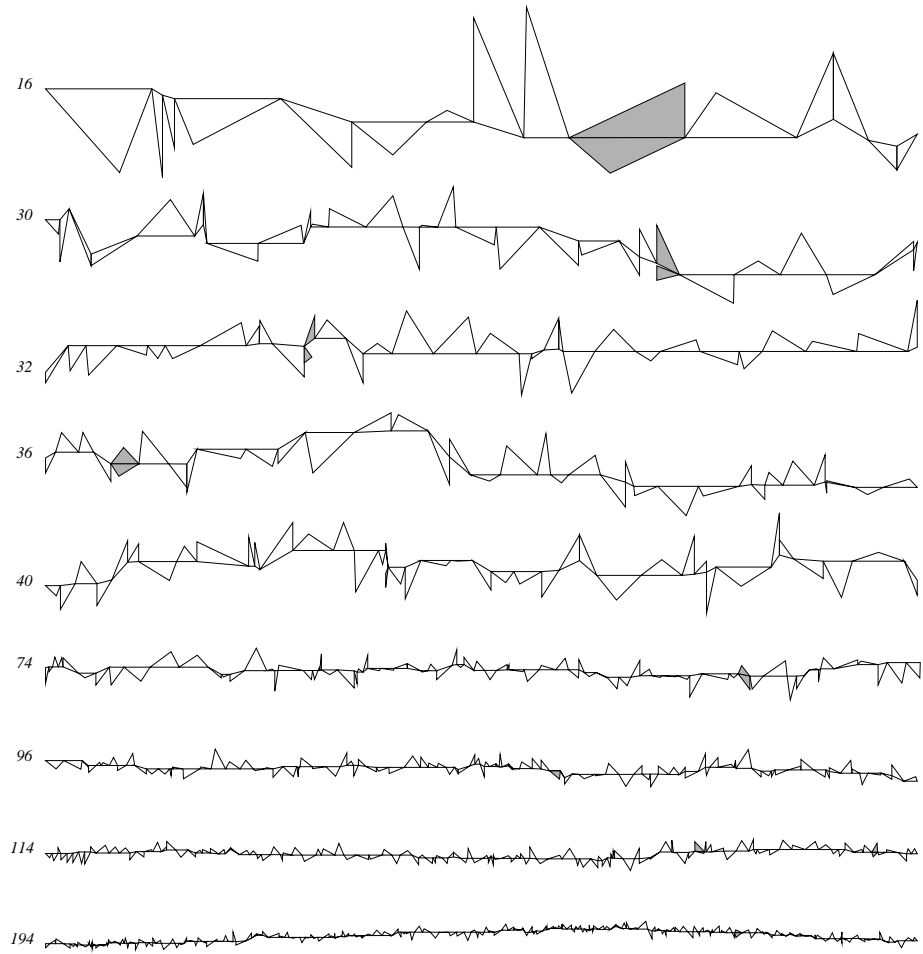}
\caption{Unfoldings of 
random convex polyhedra 
(generated by code 
from~\protect\cite{o-cgc-98}).
The number of triangles is indicated to the left of each
unfolding.
Quadrilaterals are shaded.}
\figlab{vunf.all}
\end{figure}

The next section shows that 
for polyhedra of genus zero, quadrilaterals can be avoided entirely.

\section{Genus-Zero Facet Cycles}
\seclab{facet.cycle}

\begin{theorem}
The lattice graph of any simplicial polyhedron $P$ of genus zero
contains a facet cycle $C(P)$.
\theolab{facet.cycle}
\end{theorem}
\noindent
A \emph{facet cycle} is a facet path (cf. Sec.~\secref{Overview})
that is also a cycle.

\begin{pf}
Note that any facet path in which each vertex is incident
to an even number
of path edges is a facet cycle, because it supports an Eulerian tour.

The proof is by induction on the number of polyhedron edges.
If $P$ is a tetrahedron, then there is a facet cycle,
as shown in 
Fig.~\figref{K4}.
\begin{figure}[htbp]
\centering
\includegraphics[width=0.3\linewidth]{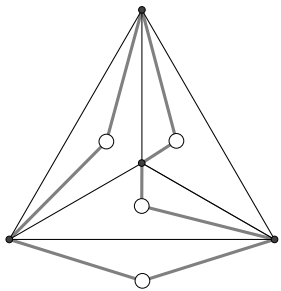}
\caption{The lattice graph of a tetrahedron contains a facet cycle.}
\figlab{K4}
\end{figure}

Otherwise, let $xy$ be any edge of $P$, and contract
it to form the lattice graph of a polyhedron $Q$.
(In a 
triangulated planar graph, at least one edge incident to any vertex can be 
contracted, unless the graph is $K_4$, which is the induction base.)
By induction, there is a facet cycle $C(Q)$ in the graph for $Q$.
See Fig.~\figref{facet.cycle}(a-c).
\begin{figure}[htbp]
\centering
\includegraphics[width=0.9\linewidth]{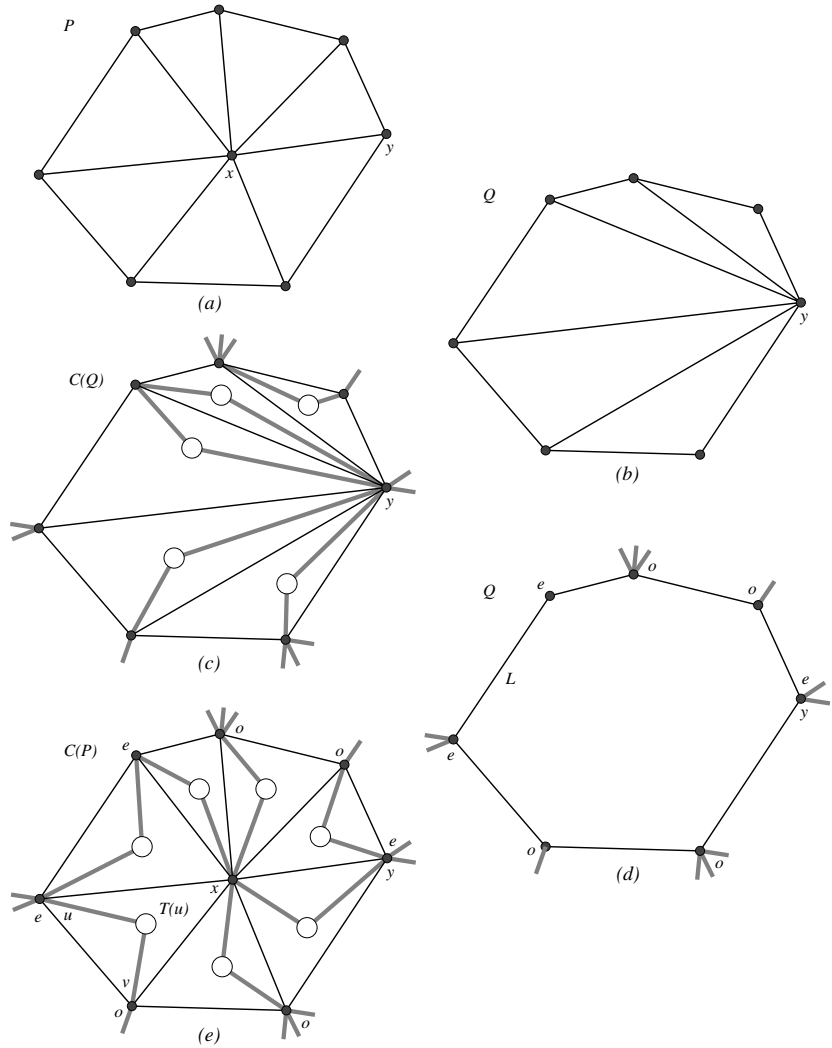}
\caption{(a) $P$; (b) $Q$ after contracting $xy$;
(c) a facet cycle on $Q$; (d) after removing edges inside
the contracted neighborhood; $e$=even, $o$=odd; (d) a facet cycle on $P$.}
\figlab{facet.cycle}
\end{figure}

Let $L$ be the link of $x$ on $P$: the edges opposite $x$ of
all triangles incident to $x$.  $L$ forms a cycle on $P$,
and on $Q$.
Now, remove the portion of $C(Q)$ that is inside $L$
on $Q$,
as in Fig.~\figref{facet.cycle}(d);
let $H$ be the resulting subgraph of $C(Q)$.
The task now is to augment $H$ on $P$ so that it is connected,
its vertices are even, and it covers all the triangles inside $L$
on $P$.

Let $u$ be a vertex of $L$.
Label $u$ \emph{odd} or \emph{even} if the number of edges of
$H$ incident to $u$ is odd or even respectively
(see Fig.~\figref{facet.cycle}(d)).
Let $T(u)$ be the triangle $\triangle u x v$ of $P$,
where $u$, $x$, and $v$ are in counterclockwise order.
We use the following rule to augment $H$.
For each $u \in L$, 
if $u$ is odd, add $(v, T(u), x)$;
if $u$ is even, add $(u, T(u), v)$.
See Fig.~\figref{facet.cycle}(e).
Call the augmented graph $C(P)$.

By construction, each triangle inside $L$ is covered by $C(P)$.
Each vertex $u \in L$ becomes incident to an added edge
from the clockwise previous $u'$, regardless of whether $u'$ is odd
or even.
If $u$ is odd, then $u$ only receives one new edge from the
clockwise previous vertex; if $u$ is even, it in addition adds
an edge to cover $T(u)$.  Consequently, every vertex on $L$
has even degree in $C(P)$.
The vertex $x$ is even because it receives an edge from
every odd vertex on $L$, and there are an even number of
odd vertices in $H$ (as in any graph), all of which are on $L$.

Next we argue that the vertices on the link $L$ are connected.
If there is an odd vertex $u$, then the added edges form a 
collection of ``arms'' extending from each odd vertex, clockwise through 
consecutive even vertices, and then connecting through $x$.  
If all vertices 
are even, then the added vertices form a cycle excluding $x$.
In either case, the vertices of $L$ become connected in $C(P)$.

Finally we show that $C(P)$ is connected.
Let $u \in L$ and let $z$ be a vertex of $C(P)$ outside of $L$.
Because $C(Q)$ is connected, there must be some path $p$
in $C(Q)$ from $z$ to a vertex on $L$; let $w \in L$ be the first such.
(Note that $L$ contains at least one triangle, and $C(Q)$ visits
two of its vertices, at least one of which is on $L$.)
The portion of the path $p$ from $z$ to $w$ is unchanged on $P$.
Because the vertices on $L$ are connected in $C(P)$, there is 
a connection from $z$ to $w$ to $u$ in $C(P)$. Thus
$C(P)$ is connected.

We have established that $C(P)$ covers the facets, is connected,
and has even degree at each vertex.  Therefore it is a facet cycle.
\end{pf}


Following this proof,
and taking care to contract an edge incident to a low-degree
vertex~\cite{ChrEpp-TCS-91},
leads to a linear-time algorithm.
Through the vertical-strip layout, this theorem permits the surface
of any triangulated genus-zero polyhedron to be unfolded to a
string of triangles joined at vertices:
\begin{cor}
The surface of any simplicial polyhedron $P$ of genus zero
may be vertex-unfolded (in linear time) into parallel strips each
containing one triangle of $P$.
\end{cor}
Theorem~\theoref{facet.cycle}
also yields an ``ideal rendering'' of any such surface
on a computer graphics system with a $1$-vertex cache:
each triangle shares one vertex with the previous triangle in
the graphics pipeline.  It is known that sharing two
vertices is not always achievable: some triangulations
do not admit a ``sequential triangulation,''
that is, an ordering of the triangles corresponding to a Hamiltonian
path in the dual graph~\cite{ahms-htfr-96}.

The restriction to simplicial polyhedra 
in Theorem~\theoref{facet.cycle} 
(and indeed in Theorems~\theoref{disk}-\theoref{any.genus} as well) 
is necessary,
for the truncated cube has no facet path:
no pair of its
eight triangles can
be adjacent in a path, but the six octagons are
not enough to separate the triangles.  

\section{Discussion}
\seclab{Discussion}
Our work raises three new open problems:
\begin{enumerate}
\item 
Does every lattice graph of a simplicial polyhedron
of genus more than zero
have a facet path,
i.e., can Theorem~\theoref{facet.cycle}
be extended to higher-genus polyhedra?

\item Does every polyhedron with faces homeomorphic to a disk
have a nonoverlapping vertex-unfolding?
The strip construction fails for faces with more than three sides.
If faces are permitted to have holes, then there are examples
that cannot be vertex-unfolded, e.g.,
a box-on-top-of-a-box (cf.~Fig.~7 of~\cite{bddloorw-uscop-98}).

\item Does every $4$-polytope,
or more generally, every polyhedral complex in
$\R^4$, have a vertex-unfolding?
It was this question that prompted our investigation.
\end{enumerate}

\paragraph{Acknowledgments}
We thank Anna Lubiw for a clarifying discussion,
and Allison Baird, Dessislava Michaylova, and Amanda Toop
for assisting with the implementation.

\bibliographystyle{alpha}
\bibliography{vunf}

\end{document}